\documentclass[showpacs]{revtex4}
\usepackage{epsfig}
\def\beq{\begin{equation}}
\def\enq{\end{equation}}
\def\beqa{\begin{eqnarray}}
\def\enqa{\end{eqnarray}}

\def\GeV{\nobreak\,\mbox{GeV}}

\def\qq{\lag\bar{q}q\rag}

\def\mix{\lag\bar{q}g\si.Gq\rag}

\def\Gd{\lag g^2G^2\rag}
\def\G3{\lag g^3G^3\rag}

\def\rh{\rho}
\def\si{\sigma}

\def\al{\alpha}
\def\be{\beta}
\def\alma{\alpha_{max}}
\def\almi{\alpha_{min}}
\def\bemi{\beta_{min}}
\def\lb{\label}
\def\nn{\nonumber}

\newcommand{\rag}{\rangle}
\newcommand{\lag}{\langle}

\begin{document}

\title{\sc
The meson $Z^+(4430)$ as a tetraquark state}

\author{M.E. Bracco}
\affiliation{Instituto de F\'{\i}sica, Universidade do Estado do Rio 
de Janeiro, Rua S\~ao Francisco Xavier 524, 20550-900 Rio de Janeiro, RJ, 
Brazil}
\email{bracco@uerj.br}
\author{S.H. Lee}
\email{suhoung@phya.yonsei.ac.kr}
\affiliation{Institute of Physics and Applied Physics, Yonsei University,
Seoul 120-749, Korea}
\author{M. Nielsen}
\email{mnielsen@if.usp.br}
\affiliation{Instituto de F\'{\i}sica, Universidade de S\~{a}o Paulo,
C.P. 66318, 05389-970 S\~{a}o Paulo, SP, Brazil}
\author{R.~Rodrigues da Silva}
\email{romulo@df.ufcg.edu.br}
\affiliation{Unidade Acad\^emica de F\'\i sica, Universidade Federal de 
Campina Grande, 58.109-900 Campina Grande, PB, Brazil}

\begin{abstract}
We test the validity of the QCD  sum rules applied to the meson $Z^+(4430)$,
by considering a diquark-antidiquark type of current with $J^{P}=0^{-}$ 
and with $J^{P}=1^{-}$. We find that, with the studied currents, 
it is possible to find an acceptable Borel window. In such a  Borel 
window we have simultaneously  a good OPE convergence and 
a pole contribution which is bigger than the continuum contribution.
We get $m_Z=(4.52\pm0.09)~\GeV$ and $m_Z=(4.84\pm0.14)~\GeV$ for the
currents with $J^{P}=0^{-}$ and $J^{P}=1^{-}$ respectively. We conclude that
the QCD sum rules results favors $J^{P}=0^{-}$ quantum numbers 
for the $Z^+(4430)$ meson.
\end{abstract}

\pacs{ 11.55.Hx, 12.38.Lg , 12.39.-x}
\maketitle


During the past years, a series of exotic charmonium like mesons, called
$X$, $Y$ and $Z$, have been discovered in $B$ mesons decays. Among them,
the charged resonance state $Z^+(4430)$, observed by Belle Collaboration
\cite{belle} in the $Z^+\to \psi^\prime\pi^+$ decay mode, is the most
intriguing one since it can not be described as ordinary $c\bar{c}$ meson.

The nature of the $Z^+(4430)$ meson is completely open and there
are already many theoretical interpretations about its structure
\cite{ding,rosner,maiani,meng,cheung,gers,qiao,lee,liu,Li,braaten,bugg,liu2,liu3,liu4,cardoso,ding2,mats}.
However, an intriguing possibility is the 
interpretation as tetraquark or molecular state. In ref. \cite{meng}, 
the closeness of the $Z^+(4430)$ mass to the threshold of $D^{*+}(2010)
\bar{D}_1(2420)$ lead the authors to consider the $Z^+(4430)$
as a $D^*\bar{D}_1$ molecule. This hypothesis was tested in ref.~\cite{lee}
by using the QCD sum rules approach, with a good agreement with the
experimental data. The interpretation of $Z^+(4430)$ as tetraquark state
was done in refs.~\cite{rosner,maiani,gers}.

Since $Z^+(4430)$ was observed in the $\psi'\pi^+$ channel, it is an 
isovector state with positive $G$-parity: $I^G=1^+$. However, nothing is 
known about its spin and parity quantum numbers. For a $D^*\bar{D}_1$ 
molecular state in s-wave, the allowed $J^P$ are
 $0^-,~1^-$ or $2^-$, although the $2^-$ assignment is probably suppressed
in the $B\to Z(4430)K$ decay, by the small phase space. 
In this work we use QCD sum rules (QCDSR) \cite{svz,rry,SNB}, to
study the two-point function of the state $Z^+(4430)$ considered as a
tetraquark state with $J^P=0^-$ and $J^P=1^-$.

In  previous calculations, the QCDSR approach was used to study
the $X(3872)$ by using a diquark-antidiquark current \cite{x3872},
 the $Z^+(4430)$ meson, by using a $D^*D_1$ molecular current 
\cite{lee} and the $Y$ mesons \cite{rapha} by using molecular and 
diquark-antidiquark type of currents.
In all cases a very good agreement with the experimental mass was obtained.


Let us consider first the $Z^+(4430)$ by using a diquark-antidiquark 
current with 
$J^P=0^-$ and positive $G$ parity. One can invoke simple arguments using 
constituent quark model to show why the tetraquark with the suggested 
quantum number could be stable.  In the constituent quark model, a 
multiquark exotic is expected to have some scalar diquark component in 
the color anti-triplet configuration, as this is the most attractive 
quark-quark channel.  However, when a tetraquark has $J^P=0^+$ quantum 
number, it would energetically be more favorable to decay in s-wave into 
two pseudo-scalar mesons.  In terms of the spin spin interaction, one can 
say that the attraction in the quark-antiquark configuration in the two 
pseudo-scalar mesons is  phenomenologically more than a factor 3 larger 
than that in  the two scalar quark-quark channel in the tetraquark
\cite{Lee08}.  However, when the tetraquark configuration has $J^P=0^-$ 
quantum number, at least one of the diquark could be in the attractive 
channel, while the remaining diquark is in the pseudo scalar channel.  
On the other hand,  it can not decay into  final states containing a 
pseudo-scalar meson in s-wave; hence the tetraquark could be quasi-stable.  
To test such configuration in a non-perturbative way, we are implementing 
the QCD sum rule method.

A possible current describing such state is given by:
\beq
j={i\epsilon_{abc}\epsilon_{dec}\over\sqrt{2}}[(u_a^TC\gamma_5c_b)
(\bar{d}_dC\bar{c}_e^T)-(u_a^TC c_b)(\bar{d}_d\gamma_5C\bar{c}_e^T)]\;,
\label{field}
\enq
where the index $T$ means matriz transposition, $a,~b,...$ are color 
indices and $C$ is the charge conjugation matriz. 

The QCD sum rules for the meson mass are constructed from the two-point 
correlation function:
\beq
\Pi(q)=i\int d^4x ~e^{iq.x}\lag 0
|T[j(x)j^\dagger(0)]|0\rag.
\lb{2po}
\enq

Phenomenologically, the correlator can be expressed as a dispersion integral
\beq
\Pi^{phen}(q^2)=\int ds\, {\rho^{phen}(s)\over s-q^2}\,+\,\cdots\,,
\label{phen}
\enq
where $\rho^{phen}(s)$ is the spectral density and the dots represent
subtraction terms. The spectral density is described, as usual, as a single 
sharp
pole representing the lowest resonance plus a smooth continuum representing
higher mass states:
\beqa
\rho^{phen}(s)&=&\lambda^2\delta(s-m_{Z}^2) +\rho^{cont}(s)\,,
\label{den}
\enqa
where $\lambda$ is proportional to the meson decay constant, $f_Z$, which
parametrizes the coupling of the current to the meson $Z^+$:
\beq
\lag0|j|Z^+\rag =f_Zm_Z^4=\lambda.
\label{lambda}
\enq

It is important to notice that  there is no one to one correspondence between
the current and the state, since the current
in Eq.~(\ref{field}) can be rewritten in terms of sum a over molecular type
currents, by the use of the Fierz transformation. However, the 
parameter $\lambda$, appearing in Eq.~(\ref{lambda}), gives a measure of the 
strength of the coupling between the current and the state.

We follow the prescription that the continuum contribution to the spectral 
density, $\rho^{cont}(s)$ in Eq.~(\ref{den}), vanishes bellow a certain 
continuum threshold $s_0$. Above this threshold, it is given by the result 
obtained with the OPE \cite{io1}:
\beq
\rho^{cont}(s)=\rho^{OPE}(s)\Theta(s-s_0)\;,
\enq

On the OPE side, we work at leading order in $\alpha_s$ and consider the
contributions of condensates up to dimension eight.  To keep the charm
quark mass finite, we use the momentum-space expression for the
charm quark propagator. The light quark part of the correlation function
is calculated in the coordinate-space. Then, the resulting light-quark 
part is Fourier transformed to the momentum space in $D$ dimensions and 
it is dimensionally regularized at $D=4$.
The correlation function in the OPE side can be written as:
\beq
\Pi^{OPE}(q^2)=\int_{4m_c^2}^\infty ds {\rho^{OPE}(s)\over s-q^2}
+\Pi^{mix\qq}(q^2)\;,
\lb{ope}
\enq
where $\rho^{OPE}(s)$ is given by the imaginary part of the
correlation function: $\pi \rho^{OPE}(s)=\mbox{Im}[\Pi^{OPE}(s)]$.

After equating the two representations of the correlation function,
assuming quark-hadron duality, making a Borel transform to both sides, and
transferring the continuum contribution to the OPE side, the sum rule
for the pseudoscalar meson $Z^+$, up to dimension-eight condensates, is 
given by:
\beq \lambda^2e^{-m_Z^2/M^2}=\int_{4m_c^2}^{s_0}ds~
e^{-s/M^2}~\rho^{OPE}(s)\; +\Pi^{mix\qq}(M^2)\;, \lb{sr}
\enq
where
\beq
\rho^{OPE}(s)=\rho^{pert}(s)+\rh^{\qq}(s)+\rh^{\lag G^2\rag}(s)
+\rh^{mix}(s)+\rh^{\qq^2}(s)\;,
\lb{rhoeq}
\enq
with
\beqa\label{eq:pert}
&&\rho^{pert}(s)={1\over 2^{9} \pi^6}\int\limits_{\almi}^{\alma}
{d\al\over\alpha^3}
\int\limits_{\bemi}^{1-\al}{d\be\over\be^3}(1-\al-\be)
\left[(\al+\be)m_c^2-\al\be s\right]^4,
\nn\\
&&\rho^{\qq}(s)=0,
\nn\\
&&\rho^{\lag G^2\rag}(s)={\Gd\over2^{8}\pi^6}
\int\limits_{\almi}^{\alma} {d\al\over\al^2}
\int\limits_{\bemi}^{1-\al}d\be\left[(\al+\be)m_c^2-\al\be s
\right]\left(m_c^2{1-\al-\be\over3\al}+{(\al+\be)m_c^2-\al\be s\over4\be}
\right)
\nn\\
&&\rho^{mix}(s)=0,
\nn\\
&&\rho^{\qq^2}(s)=-{m_c^2\qq^2\over 12\pi^2}\sqrt{1-4m_c^2/s},
\enqa
\beqa
\Pi^{mix\qq}(M^2)={m_c^2\mix\qq\over 24\pi^2}\int_0^1
d\al\,{e^{-m_c^2
\over\al(1-\al)M^2}\over1-\al}\bigg[{m_c^2\over\al M^2}-\al\bigg]\,.
\label{8m2}
\enqa
The integration limits are given by $\almi=({1-\sqrt{1-
4m_c^2/s})/2}$, $\alma=({1+\sqrt{1-4m_c^2/s})/2}$ and $\bemi={\al
m_c^2/( s\al-m_c^2)}$.

One should note that a evaluation of the higher dimension 
condensate contributions is technically difficult and non-trivial, 
which cannot be obtained by a simple routine iteration of the 
quark propagator in an external field. Violation of the factorization 
hypothesis become increasingly important in higher dimensions and so 
the results become increasingly model dependent, as more condensates 
will have to be introduced if factorization is not valid \cite{bnp}.

Similarly to the results in ref.~\cite{lee}, the current in 
Eq.~(\ref{field}) does not get contribution from the quark and mixed 
condensates. This is very 
different from the OPE behavior obtained for the diquark-antidiquark current
used for the $X(3872)$ and $Y(4660)$ mesons in refs.~\cite{x3872,rapha}, 
but very similar to the OPE
behavior obtained for the axial double-charmed meson $T_{cc}$, also
described by a diquark-antidiquark current \cite{tcc}.

In the numerical analysis, the input values are taken as \cite{SNB,narpdg}:
$m_c(m_c)=(1.23\pm 0.05)\,\GeV $,
$\lag\bar{q}q\rag=\,-(0.23\pm0.03)^3\,\GeV^3$,
$\lag\bar{q}g\si.Gq\rag=m_0^2\lag\bar{q}q\rag$ with $m_0^2=0.8\,\GeV^2$,
$\lag g^2G^2\rag=0.88~\GeV^4$.

\begin{figure}[h]
\centerline{\epsfig{figure=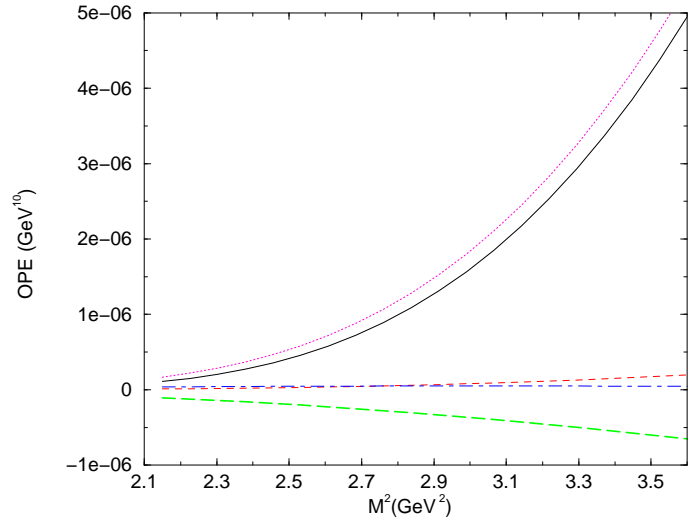,height=70mm}}
\caption{The OPE convergence in the region $2.2 \leq M^2 \leq
3.5~\GeV^2$ for $\sqrt{s_0} = 4.9$ GeV. Perturbative
contribution (dotted line), $\langle g^2G^2\rangle$ contribution
(dashed line), $\langle \bar{q}q\rangle^2$ contribution (lon-dashed line),
$\mix\qq$ (dot-dashed line) and the total contribution (solid line).}
\label{figconvtcc}
\end{figure}

We evaluate the sum rule in the Borel range $2.2 \leq M^2 \leq 3.5\GeV^2$.
To determine the allowed Borel window, we analyse the OPE convergence and 
the pole contribution: the minimum value of the Borel mass is fixed by 
considering the convergence of the OPE, and the maximum value of the Borel 
mass is determined by imposing that the pole contribution must be bigger 
than the continuum contribution. To fix the continuum
threshold range we extract the mass from the sum rule, for a given $s_0$,
and accept such value of $s_0$ if the obtained mass is around 0.5 GeV 
smaller than $\sqrt{s_0}$. However, in this case, to be able to compare
our results with the results obtained by using a molecular type current
in ref.~\cite{lee}, we use the same continuum range as in ref.~\cite{lee}:
$4.8\leq \sqrt{s_0} \leq5.0$ GeV.

From Fig.~\ref{figconvtcc} we see that we obtain a quite good OPE
convergence for $M^2\geq 2.3$ GeV$^2$. Therefore, we  fix the lower
value of $M^2$ in the Borel window as $M^2_{min} = 2.3$ GeV$^2$.
This figure also shows that the dimension-eight condensate contribution 
is very small as compared with the four-quark condensate contribution.

\begin{figure}[h]
\centerline{\epsfig{figure=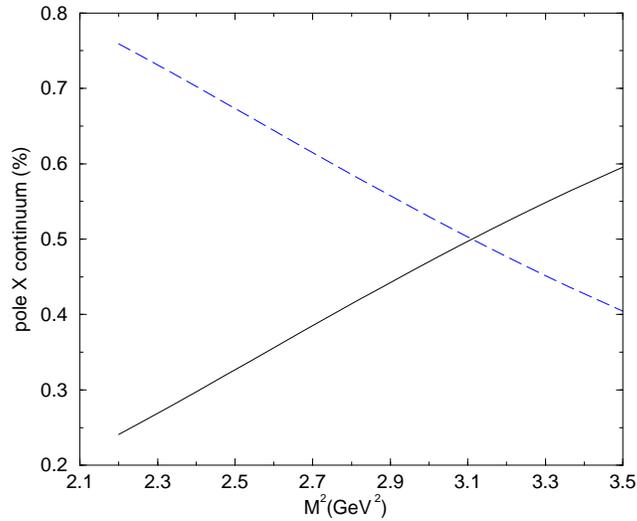,height=70mm}}
\caption{The dashed line shows the relative pole contribution (the
pole contribution divided by the total, pole plus continuum,
contribution) and the solid line shows the relative continuum
contribution for $\sqrt{s_0}=4.9~\GeV$.}
\label{figpvc}
\end{figure}

The comparison between pole and
continuum contributions for $\sqrt{s_0} = 4.9$ GeV is shown in
Fig.~\ref{figpvc}, from where  we see that the pole contribution
is bigger than the continuum for $M^2\leq3.1~\GeV^2$. The same analysis 
for the other values of the
continuum threshold gives $M^2 \leq 2.9$  GeV$^2$ for $\sqrt{s_0} = 
4.8~\GeV$ and $M^2 \leq 3.3$  GeV$^2$ for $\sqrt{s_0} = 5.0~\GeV$.

\begin{figure}[h]
\centerline{\epsfig{figure=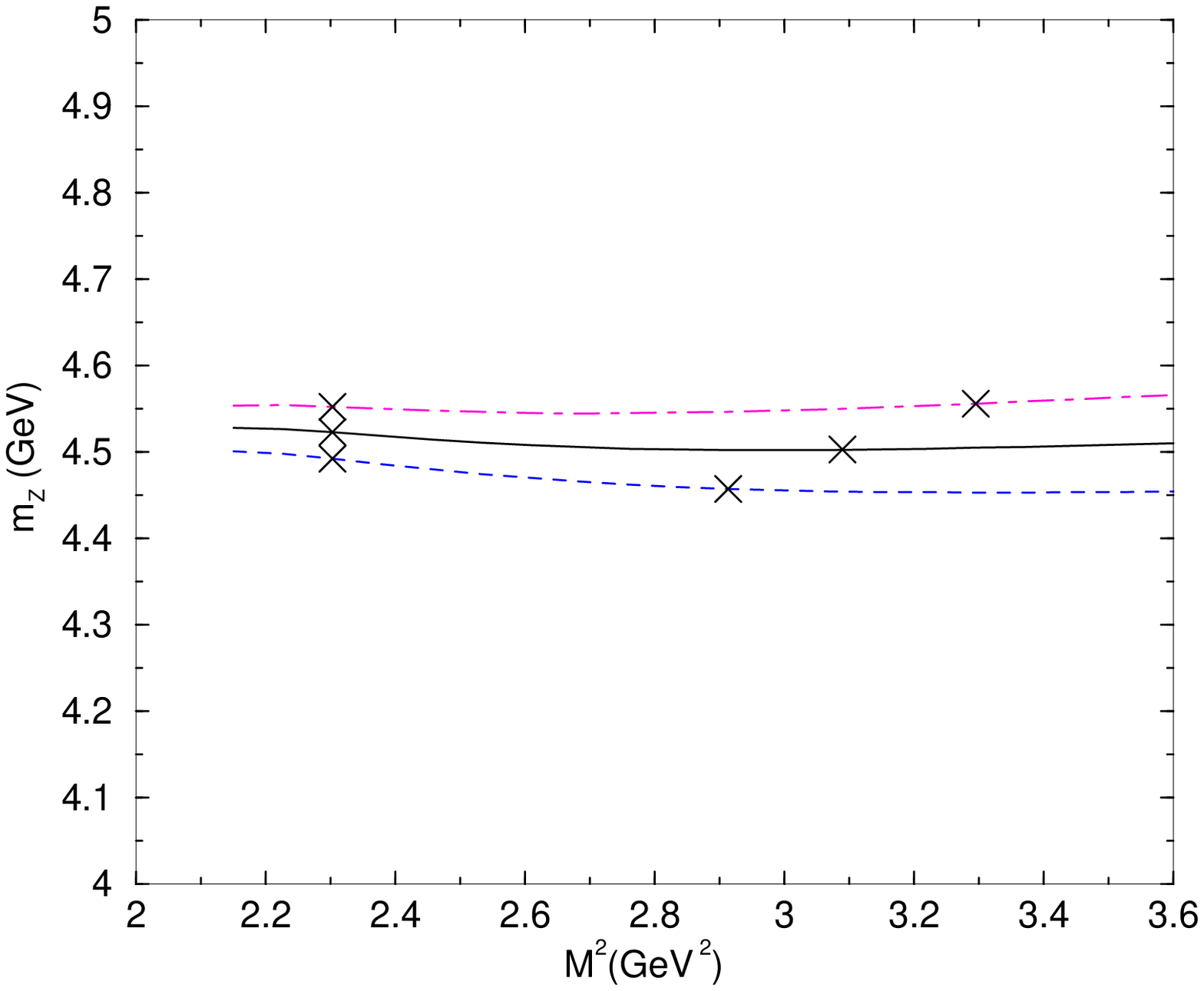,height=70mm}}
\caption{The $Z^+$ with $J^P=0^-$ meson mass as a function of the sum 
rule parameter
($M^2$) for different values of $\sqrt{s_0}$: $\sqrt{s_0}=4.8~\GeV$ dashed 
line,$\sqrt{s_0}=4.9~\GeV$ solid line and $\sqrt{s_0}=5.0~\GeV$ dot-dashed 
line. The crosses
indicate the region allowed for the sum rules.}
\label{figmx}
\end{figure}

To extract the mass $m_Z$ we take the derivative of Eq.~(\ref{sr})
with respect to $1/M^2$, and divide the result by Eq.~(\ref{sr}).
In Fig.~\ref{figmx}, we show the $Z^+$ meson mass, for different values
of $\sqrt{s_0}$, in the relevant sum rule window, with the upper and lower 
validity limits indicated.  From this figure we see that
the results are very stable as a function of $M^2$.

To check the dependence of our results with the value of the
charm quark mass, we fix $\sqrt{s_0}=4.9~\GeV$ and vary the charm quark mass
in the range $m_c=(1.23\pm0.05)~\GeV$. Using $2.5\leq M^2\leq 3.1~\GeV^2$
we get: $m_{Z} = (4.51\pm0.06)~\GeV$. Including the uncetainty due to the
value of the continuum threshold and the value of the Borel parameter we 
arrive at
\beq
m_{Z_{(0^-)}} = (4.52\pm0.09)~\GeV,
\label{massz0}
\enq
which is a little bigger than the experimental value \cite{belle}, but still 
consistent with it, considering the uncertanties. Comparing our result with
the result obtained in ref.~\cite{lee}: $m_{D^*D_1}=(4.40\pm0.10)
\GeV$, where the $Z^+(4430)$ was considered by using a $D^*D_1$ molecular 
current with $J^P=0^-$, we see that the result in ref.~\cite{lee} is in a 
better 
agreement with the experimental value. However, as mentioned above, since 
there is no one to one corresponde between the structure of the current 
and the state, we can not use this result to conclude that the $Z^+(4430)$ 
is better explained as a molecular state than as a diquark-antidiquark 
state. To
get a measure of the coupling between the state and the current, we use
Eq.~(\ref{sr}) to evaluate the parameter $\lambda$, defined in 
Eq.(\ref{lambda}). We get:
\beq
\lambda_{Z_{(0^-)}} = \left(3.75\pm0.48\right)\times 10^{-2}~\GeV^5,
\label{ladi}
\enq
while for the current used in ref.~\cite{lee} we get:
\beq
\lambda_{D^*D_1} = \left(5.66\pm1.26\right)\times 10^{-2}~\GeV^5.
\label{lamo}
\enq
Therefore, it is possible to conclude that the physical particle with
$J^P=0^-$ and quark content $c\bar{c}u\bar{d}$ couples with a larger
strength with the molecular $D^*D_1$ type current than with the
current in Eq.(\ref{field}).

We now consider the $Z^+(4430)$ by using a diquark-antidiquark current with 
$J^P=1^-$ and positive $G$ parity. The lowest-dimension interpolating 
operator describing such current is given by:
\beq
j_\mu={\epsilon_{abc}\epsilon_{dec}\over\sqrt{2}}[(u_a^TC\gamma_5c_b)
(\bar{d}_d\gamma_\mu\gamma_5 C\bar{c}_e^T)+(u_a^TC\gamma_5\gamma_\mu c_b)
(\bar{d}_d\gamma_5C\bar{c}_e^T)]\;.
\label{field2}
\enq
The two-point correlation function is now given by:
\beq
\Pi_{\mu\nu}(q)=i\int d^4x ~e^{iq.x}\lag 0
|T[j_\mu(x)j^\dagger_\nu(0)]
|0\rag=-\Pi(q^2)(g_{\mu\nu}q^2-q_\mu q_\nu),
\lb{2po2}
\enq
from where we get 
\beq
\Pi_{\mu}^{\mu}(q)=-3q^2\Pi(q^2),
\lb{inv}
\enq
and, therefore, we can write a sum rule for $\Pi(q^2)$ as before.
The spectral density is now given by
\beqa\label{rhoope}
&&\rho^{pert}(s)=-{1\over 2^{8}3 \pi^6s}\int\limits_{\almi}^{\alma}
{d\al\over\alpha^3}
\int\limits_{\bemi}^{1-\al}{d\be\over\be^3}(1-\al-\be)
\left[(\al+\be)m_c^2-\al\be s\right]^3\left[m_c^2-2m_c^2(\alpha+\beta)+
\al\be s\right],
\nn\\
&&\rho^{\qq}(s)=0,
\nn\\
&&\rho^{\lag G^2\rag}(s)={m_c^2\Gd\over3^22^{9}\pi^6s}
\int\limits_{\almi}^{\alma} {d\al}
\int\limits_{\bemi}^{1-\al}{d\be\over\beta^3}(1-\al-\be)\left[4(2\al+2\be-1)
m_c^2-{3m_c^2\beta\over\alpha}-\be s(7\alpha-3)\right],
\nn\\
&&\rho^{mix}(s)={m_c\mix\over2^{6}3 \pi^4s}\int\limits_{\almi}^{\alma}
{d\al\over\alpha}
\int\limits_{\bemi}^{1-\al}{d\be\over\be^2}(2\alpha+\beta)\left[(\al+\be)
m_c^2-\al\be s\right],
\nn\\
&&\rho^{\qq^2}(s)=-{\qq^2\over 36\pi^2}\left({5m_c^2\over s}-{1\over2}
\right)\sqrt{1-4m_c^2/s},\nn\\
&&\rho^{mix\qq^2}(s)=-{\qq\mix\over3^22^4\pi^2s}(1+4m_c^2/s)
\sqrt{1-4m_c^2/s},
\enqa

\beqa
&&\Pi^{mix\qq}(M^2)=-{\qq\mix\over3^22^4 \pi^2}\left({2\over3}-3\int_0^1
d\al\,\exp\!\left[{-{m_c^2
\over\al(1-\al)M^2}}\right]\bigg[\al-2\alpha^2+{2m_c^2
\over M^2}\bigg]\right)\,.
\label{8m2}
\enqa

\begin{figure}[h]
\centerline{\epsfig{figure=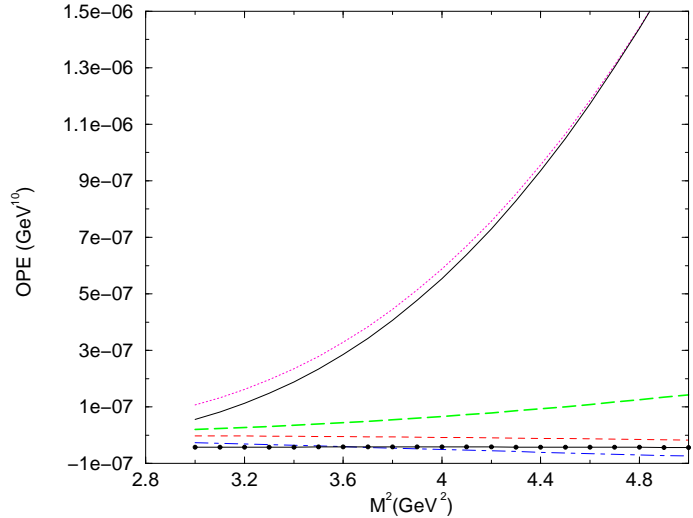,height=70mm}}
\caption{The OPE convergence for the sum rule for $Z^+$ with $J^{P}=1^-$, 
using $\sqrt{s_0} =5.3$ GeV. The dotted, dashed, long-dashed, 
dot-dashed, solid with dots and solid lines give, respectively, the 
perturbative, gluon condensate, mixed condensate, four-quark condensate,
dimension-eight condensate and total contributions.}
\label{figope}
\end{figure}

Although with this current we still do not get
contribution from the quark condensate, we do get contribution from
the mixed condensate. As can be seen by  Fig.~\ref{figope}, the mixed 
condensate contribution is of the same order as the four-quark condensate
contribution, but with opposite signal. The contribution of the 
dimension-eight condensate is now of the same order as the four-quark
condensate contribution, for small values of $M^2$.

In this case we find that the continuum threshold is in the range
$\sqrt{s_0}=(5.3\pm0.1)~\GeV$ and, from Fig.~\ref{figope}, we see
that there is a good OPE convergence for $M^2\geq3.9~\GeV^2$. 

The upper limits for $M^2$ for each value of $\sqrt{s_0}$ are given in 
Table I, from where we see that the Borel window in this case has higher
values of the Borel parameter, as compared with the case for $Z^+$ with 
$J^P=0^-$.

\begin{center}
\small{{\bf Table I:} Upper limits in the Borel window for $Z^+$ with 
$J^P=1^-$.}
\\
\begin{tabular}{|c|c|}  \hline
$\sqrt{s_0}~(\GeV)$ & $M^2_{max}(\GeV^2)$  \\
\hline
 5.2 & 4.4 \\
\hline
5.3 & 4.7 \\
\hline
5.4 & 5.0 \\
\hline
\end{tabular}\end{center}

\begin{figure}[h]
\centerline{\epsfig{figure=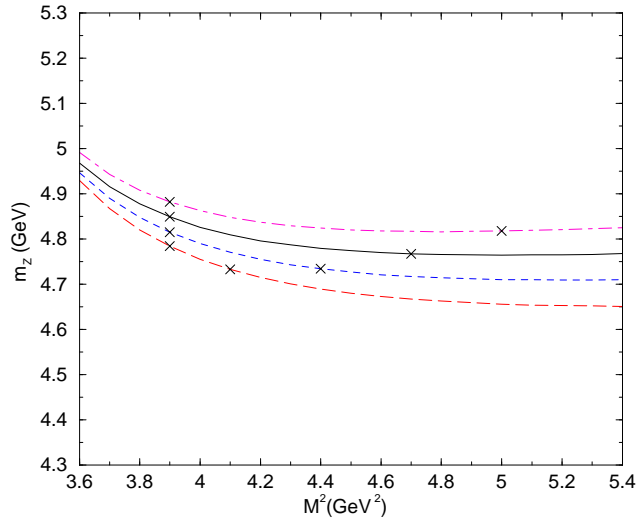,height=70mm}}
\caption{The $Z^+$ with  $J^P=1^-$ meson mass as a function of the sum rule 
parameter for different values of $\sqrt{s_0}$: $\sqrt{s_0}=5.1~\GeV$ 
long-dashed line, $\sqrt{s_0}=5.2~\GeV$ dashed 
line, $\sqrt{s_0}=5.3~\GeV$ solid line and $\sqrt{s_0}=5.4~\GeV$ dot-dashed 
line. The crosses indicate the region allowed for the sum rules.}
\label{figmz}
\end{figure}

In the case of $Z^+$ with  $J^P=1^-$ we get a worse Borel stability than 
for the $Z^+$ with $J^P=0^-$, in the allowed sum rule window, as a function 
of $M^2$, as can be seen by  Fig.~\ref{figmz}. We also observe that the 
results are, in this case, more sensitive to the values of $m_c$.

Using the Borel window, for each value of $s_0$, to evaluate the mass, 
and then varying the value of the continuum threshold in the
range $5.2\leq \sqrt{s_0} \leq5.4$
GeV, we get $m_{Z_{(1^-)}}= (4.80\pm0.08)~\GeV$.

Because of the complex spectrum of the exotic states, some times lower 
continuum threshold values are favorable in order to completely eliminate the
continuum above the resonance state. Therefore, in Fig.~\ref{figmz} we
also include the result for $\sqrt{s_0}=5.1\GeV$.
We see that we get a very narrow Borel window, and for values of the 
continuum threshold smaller than 5.1 GeV there is no allowed Borel window. 
Taking into account the variations on $M^2$, $s_0$ and $m_c$ in
the regions indicated above we get:
\beq
\lb{massz1}
m_{Z_{(1^-)}}=  (4.84\pm0.14)~\GeV~,
\enq
which is much bigger than the experimental value and bigger than the 
result obtained using the current with $J^P=0^-$ in Eq.~(\ref{massz0}).

For the value of the parameter $\lambda$ defined in Eq.~(\ref{lambda}) we 
get:
\beq
\lambda_{Z_{(1^-)}} = \left(8.36\pm0.85\right)\times 10^{-5}~\GeV^5.
\label{la2}
\enq


In conclusion,
we have presented a QCDSR analysis of the two-point
function of the recently observed $Z^+(4430)$ meson, considered as a 
tetraquark state, with a diquark-antidiquark configuration. Since the
spin-parity quantum numbers of the $Z^+(4430)$ meson are not known, we
have considered two different possibilities: $J^P=0^-$ and
$J^P=1^-$. We have found a very good OPE convergence for these two cases, 
although this is not in general the case for tetraquark 
states \cite{tetra}. We got a $Z^+$ mass in some agreement with the 
experimental result in the case with $J^P=0^-$. However, in the case 
$J^P=1^-$,
we got a much higher  value for the mass. This is consistent with the
expectation from the constituent quark model, since in this model the scalar 
diquark component in the color anti-triplet configuration is the most 
attractive quark-quark channel.

Comparing our result, for the case $J^P=0^-$, with the case where the 
$Z^+(4430)$ meson was considered by using a $D^*D_1$ molecular current, 
also with $J^P=0^-$ \cite{lee}, the differences are also not really big. 
Since 
there is no one to one corresponde between the structure of the current 
and the state, we can not conclude that the $Z^+(4430)$ 
is better explained as a molecular state than as a diquark-antidiquark state.
However, comparing the results obtained for the quantum numbers
$J^P=0^-$ and $1^-$, from our calculations we  conclude that
the $Z^+(4430)$ is probably a $J^P=0^-$ state.

\section*{Acknowledgements}
{This work has been partly supported by FAPESP and CNPq-Brazil,
and by the Korea Research Foundation KRF-2006-C00011.}

\end{document}